\documentclass[aps,groupedaddress,superscriptaddress,amsmath,amssymb,twocolumn,prb]{revtex4-2}
\usepackage{bm}
\usepackage[pdftex]{graphicx}
\usepackage{xcolor}
\usepackage{color}
\usepackage[normalem]{ulem}
\usepackage{enumerate}
\usepackage{epstopdf}
\usepackage{braket}
\usepackage{hyperref}
\usepackage{caption}
\usepackage{blindtext}

\newcommand*\mean[1]{\overline{#1}}

\begin{document}

\title{Analytical approach to the magneto-fluorescence of triplet excitons}

\author{Yan Sun}
\affiliation{LPS, Universit\'e Paris-Saclay, CNRS, UMR 8502, F-91405 Orsay, France}
\author{A.D. Chepelianskii}
\affiliation{LPS, Universit\'e Paris-Saclay, CNRS, UMR 8502, F-91405 Orsay, France}

\begin{abstract}
The fluorescence of triplet excitons and color-centers is strongly dependent on magnetic field that mixes the zero field spin eigenstates that determine the radiative recombination rates back into the singlet ground state through spin-orbit coupling. For films of molecules, and polycrystalline color-centers samples an average over molecular orientations has to be performed to model the magneto-fluorescence lineshapes. This limits our analytical understanding of the lineshapes and complicates the analysis of the fluorescence dependence on magnetic field. Here, we present a framework that allows to average over triplet molecular orientations analytically. Our approach provides very accurate numerical routines computing precisely the averages matrix elements that appear in magneto-fluorescence and semi-analytical approximations that can be used to model experimental traces.
\end{abstract}


\maketitle

\section{Introduction}

Spin dependent fluorescence is central to many quantum sensing techniques that rely on ground-state or excited-triplet spin manifolds \cite{awschalom2018quantum,wolfowicz2021quantum,doherty2013nitrogen}. The fluorescence yield is governed by the overlap of the spin‐wavefunction with the zero‐magnetic-field eigenstates that emerge via spin–orbit–mediated intersystem crossing \cite{bayliss2018site,sun2025spin,robledo2011spin}. Furthermore, the magnetic‐field dependence of the fluorescence lineshape is controlled by the relative orientation between the applied field and the triplet fine‐structure axes, which has been seen in diverse systems such as defects in solids like NV centers in diamond \cite{tetienne2012magnetic,barry2020sensitivity, veys2023modeling} , triplet excitons in organics \cite{oxborrow2012room,bayliss2014geminate,sun2024cascade,mena2024room} and more recently organic di-radicals \cite{ai2018efficient,chowdhury2025bright}. In systems comprising randomly oriented triplets, the observed lineshape thus corresponds to an ensemble average over orientations, which is often computed using Monte Carlo simulations \cite{scheidler1996monte,yatskou2001nonisotropic,bobbert2007bipolaron,andrews2004using}. However, relying solely on Monte Carlo averaging makes it difficult to extract analytical insights into the lineshape behavior or to perform a precise theory–experiment comparison, since fitting simulated data to measurements can become time-consuming task. 

Here, we introduce an analytical framework for computing the ensemble average of the matrix elements relevant to spin-dependent fluorescence. We develop both a series representation and asymptotic approximations that accurately reproduce the low‐field and high‐field limits while retaining good accuracy throughout the intermediate magnetic‐field regime. For this purpose we introduce a theoretical method to analytically incorporate the effect of magnetic field orientation. The proposed approach reduces computational effort while maintaining fidelity to the physical behavior of triplet excitons. Moreover, it provides a transparent analytic connection between field orientation effects and experimentally observed MPL spectra. The framework presented here offers a new approach for the theoretical treatment of spin-dependent processes in excitonic and molecular systems.

\section{Theory}

In the molecular frame the Hamiltonian of a triplet in a magnetic field takes the following form:
\begin{align}
{\hat H} = D_z \left( {\hat S}_z^2 - \frac{2}{3} \right) + E({\hat S}_x^2-{\hat S}_y^2) + \mathbf{B} \cdot \mathbf{{\hat S}}
\label{S1DzBxBz}
\end{align}
where ${\hat S}_{x,y,z}$ are the spin-matricies for a spin $S=1$. The eigenvalues of this Hamiltonian are $\lambda_0, \lambda_1, \lambda_2$.

At zero magnetic field eigenvectors are given by $|Z\rangle = |0\rangle$, $|X\rangle = \frac{|1\rangle - |-1\rangle}{\sqrt{2}}$ and $|Y\rangle = \frac{|1\rangle + |-1\rangle}{\sqrt{2}}$. These states form an orthonormal basis corresponding to spin projections along the $Z$, $X$, $Y$ respectively. They obey ${\hat S^2_x}|X\rangle = {\hat S^2_y}|Y\rangle = {\hat S^2_z}|Z\rangle = 0$ and are associated with radiative recombination rates $k^0_{w,S}$ ($w=X,Y,Z$).

Omitting coherence effects in the recombination of superpositions of zero field eigenfunctions, we can write the radiative recombination rate  into the singlet ground-state for an eigenstate $|\psi_n(\mathbf{B})\rangle$ at finite $\mathbf{B}$ as: 
\begin{align}
k_{n}(\mathbf{B}) = \sum_{w=X,Y,Z} k^0_{w} |\langle \psi_n(\mathbf{B})|w\rangle|^2
\end{align}
In this expression the magnetic field dependence of the rates enters only through $|\psi_n(\mathbf{B})\rangle$ \cite{tetienne2012magnetic,el1972pmdr,budil1991chlorophyll,kraffert2014charge}. 
The fluorescence yield changes with $k_{S}(\mathbf{B})$. If the changes of the rates with magnetic field are small, it is possible to linearize the rate equations near their values at $\mathbf{B}=0$. In this case the isotropic orientation-averaged PL spectrum will be proportional to the angle average of the wave function overlaps $\mean{|\langle \psi_n(\mathbf{B})|X,Y,Z\rangle|^2}$, where the overbar denotes the isotropic angle averaging. The focus of this work is to compute these averages.

For this calculation it is convenient to make the connection between the wave function overlaps and matrix elements of the fine structure tensor:
\begin{align}
  \Sigma_{z,n} &= \mean{ \langle \psi_n | {\hat S}_z^2 | \psi_n \rangle} \\
  \Sigma_{xy,n} &= \mean{ \langle \psi_n | {\hat S}_x^2 - {\hat S}_y^2 | \psi_n \rangle}
\end{align}
The average eigenstate projectors on the zero magnetic field eigenbasis can then be written as:
\begin{align}
  \mean{ |\langle \psi_n | Z \rangle|^2 } &= 1 - \Sigma_{z,n} \\
  \mean{ |\langle \psi_n | X \rangle|^2 } &= (\Sigma_{z,n} - \Sigma_{xy,n})/2 \\
  \mean{ |\langle \psi_n | Y \rangle|^2 } &= (\Sigma_{z,n} - \Sigma_{xy,n})/2  
\end{align}

The advantage of introducing the quantities $\Sigma$ is that these values are obtained from derivatives of the eigenvalues $\lambda_n$. Thus the powder (isotropic) average of the overlaps $|\langle \psi_n(\mathbf{B})|w\rangle|^2$ are linked to isotropic averages of the eigenvalues:
\begin{align}
\Sigma_{z,n} = \frac{2}{3} + \partial_{D_z} \mean{\lambda_n} \;,\; \Sigma_{xy,n} = \partial_{E} \mean{\lambda_n}
\end{align}
Since ${\rm Tr} {\hat H} = \sum_n \lambda_n = 0$, we have the following identities that connect
\begin{align}
\sum_n \Sigma_{z,n} = 2 \;,\; \sum_n \Sigma_{xy,n} = 0
\end{align}
so knowing two of the quantities $\Sigma_{z,n}$ and $\Sigma_{xy,n}$ is sufficient, in the following we will focus on the case $n=0,2$.

\section{Expansion around average characteristic polynomial}

Random matrix theory provides elegant analytical techniques to find analytically the average eigenvalue distribution of eigenvalues \cite{beenakker1997random,mehta2004random,potters2020first}. However these techniques generally apply to the limit of large matrix dimensions which is not the case here. Computing the average resolvent operator, a central quantity in random matrix theory,  is in principle feasible. However even in the simpler case $E=0$, the resulting expression for the average resolvent becomes very complicated. Moreover even if we find some simplifications, this calculation would give the average distribution of the eigenvalues and not the average of each eigenvalue ${\mean{\lambda_n}}$, thus an alternative approach tailored for this problem is needed. Here we propose to approximate the averages $\mean{\lambda_n}$ by an expansion around the average characteristic polynomial:
\begin{align}
\mean{\det(u {\hat I} - {\hat H})} = u^3 -\left(B^2 + E^2 + D_z^2/3\right) u + \frac{2}{27} q
\end{align}
with 
\begin{align}
q = D_z^3 - 9 D_z E^2 
\end{align}

Using the Cardan formulas the average characteristic equation has three roots $u_0,u_1,u_2$, that can be expressed as function of the discriminant
\begin{align}
\Delta = (3 B^2 + D_z^2 + 3 E^2)^3 - (D_z^3 - 9 D_z E^2)^2
\end{align}
and through the following expressions (where $\omega = e^{2 \pi i/3}$) 
\begin{align}
u_2 &= \frac{1}{3}\left[ (-q + i \sqrt{\Delta})^{1/3} + (-q - i \sqrt{\Delta})^{1/3} \right]\\
u_1 &= \frac{1}{3}\left[ \omega^2 (-q + i \sqrt{\Delta})^{1/3} + \omega (-q - i \sqrt{\Delta})^{1/3} \right] \\
u_0 &= \frac{1}{3}\left[ \omega (-q + i \sqrt{\Delta})^{1/3} + \omega^2 (-q - i \sqrt{\Delta})^{1/3} \right]
\end{align}
For $D_z > 0$ the eigenvalues $u_n$ are ordered by increasing energy. The subsequent discussion will therefore focus on this regime. The expressions can be $D_z < 0$ obtained by swapping the order of the eigenvalues (ground state eigenvector is given by $\lambda_2$ and so forth).

Using the roots of the average characteristic polynomial $u_n$ the eigenvalue equation for an arbitrary $\mathbf{B}$ can be written in the following form:
\begin{align}
\det(\lambda {\hat I} - {\hat H}) = \prod_{n=0}^{2} (\lambda - u_n) - \epsilon 
\end{align}
where the parameter $\epsilon$ characterizes the fluctuations of the characteristic polynomial around its mean value.
\begin{align}
\epsilon = \frac{D_z (2 B_z^2 - B_y^2 - B_x^2)}{3} + E (B_x^2 - B_y^2)
\end{align}
clearly for an isotropic average $\mean{\epsilon} = 0$. The random direction of the magnetic field ${\mathbf B}$ only changes the offset of the characteristic polynomial. This occurs because the only term combining $\lambda$ and powers of $B$ in the characteristic polynomial comes from the Zeeman splitting for which, $\det(\lambda {\hat I} - \mathbf{S} \cdot \mathbf{B}) = \lambda(\lambda^2 - B^2)$. Due to this simple form, it is possible to use the Lagrange inversion theorem to obtain a series expansion in powers of $\epsilon$ giving the eigenvalues $\lambda_0$:
\begin{align}
\lambda_0 = u_0 + \sum_{m=1}^{\infty} \frac{\epsilon^m}{m!} \partial^{m-1}_{z=u_0} \frac{1}{(z-u_1)^m(z-u_2)^m}
\end{align}
The advantage of this series over other expansions is that the isotropic averages of the powers ${\mean \epsilon^m}$ can be easily computed. We note that this expansion is valid for any ordering of the eigenvalues $\lambda_n$ so the expansion for ${\mean \lambda_1}$ can be obtained through the permutation of $u_0$ and $u_1$, and  ${\mean \lambda_2}$ through the permutation of $u_0$ and $u_2$. In the following provide only formulas for $\mean{\lambda_0}$ as the formulas for other average eigenvalues can be obtained by simple permutation of the indices.
\begin{align}
\mean{\lambda_0} = u_0 + \sum_{m=1}^{\infty} \mean{\epsilon^m} \frac{S_n}{(u_1-u_0)^{2m-1}(u_2-u_0)^{2m-1}}
\label{eq:avlambda}
\end{align}
The Tab.~\ref{tabLagrange} provides the expressions for $S_{n}$ and $\epsilon^m$ up to order $m=4$, we computed these average to order $m=10$ using the python simpy symbolic calculation packages, and use it to provide efficient python numerical routines implanting the Lagrange summation and its derivatives. The code is available at \cite{tiso_deth}.
\begin{center}
\begin{table}
\begin{tabular}{|c|c|c|}
  \hline $n$ & $\mean{ \epsilon^m }$ & $S_n$ \\
  \hline $1$ & $0$ & $1$ \\
  \hline $2$ & $\frac{4}{45} B^4 (D_z^2 + 3 E^2)$ & $-3 u_0$ \\
  \hline $3$ & $\frac{16}{945} B^6 D_z(D_z^2 - 9 E^2)$ & $16 u_0^2 - u_1 u_2$ \\
  \hline $4$ & $\frac{16}{945} B^8 (D_z^2 + 3 E^2)^2$ & $-15 u_0 (7 u_0^2 - u_1 u_2)$ \\
  \hline
\end{tabular}
\caption{Analytical expression for the coefficients of the power series Eq.~(\ref{eq:avlambda})  }
\label{tabLagrange}
\end{table}
\end{center}

\begin{figure}[h]
\centerline{
\includegraphics[clip=true,width=9cm]{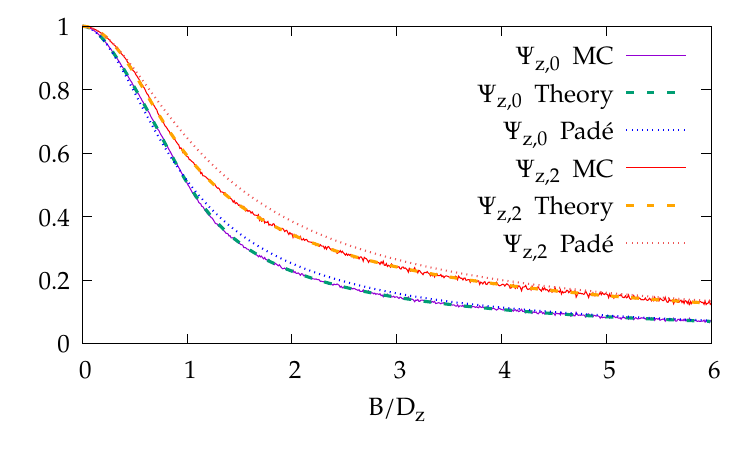} 
}
\caption{Isotropic average of $\Psi_{z,0}$ and $\Psi_{z,2}$, the normalized versions of $\Sigma_{z,0}$ and $\Sigma_{z,2}$ defined in Eqs.~(\ref{PsiZ0},\ref{PsiZ2}), for $E = 0.2 D_z$. The averages are computed using a numerical Monte-Carlo method, the series expansion around averaged characteristic polynomial and the semi-analytic Pad\'e approximation (see Tab.~\ref{tabalpha123}). The difference between MC and the series expansion is within the numerical Monte Carlo noise. The Pad\'e approximation is accurate in the limits $B \rightarrow 0$ and $B \rightarrow \infty$ with sufficient 10\% accuracy at $B \sim D_z$, the accuracy is lower in this regime because this approximation takes into account only on the second order term of the power series.}
\label{fig1}
\end{figure}

\section{Semi-analytic lineshapes, comparison with MC}

To obtain a simpler semi-analytical approximation for the elementary lineshapes $\Sigma_{z,n}, \Sigma_{xy,n}$ we normalize them to look like a positive peak of amplitude 1 at $B=0$:
\begin{align}
  &\Psi_{z,0} = -\frac{3}{2} \Sigma_{z,0}+1 = \frac{3}{2} \mean{ |\langle \psi_0 | Z \rangle|^2 } -\frac{1}{2} \label{PsiZ0} \\
  &\Psi_{z,2} =  3 \Sigma_{z,2} - 2 = - 3 \mean{ |\langle \psi_2 | Z \rangle|^2 } + 1 \label{PsiZ2}
\end{align}
The averages of ${\hat S}_x^2 - {\hat S}_y^2$, $\Sigma_{xy,2}$ already goes to unity at $B=0$ and vanishes at $B\rightarrow \infty$ and thus does not need no resealing.

To describe these lineshapes semi-analytically, we introduce a Pad\'e approximant decomposition:
\begin{align}
  P(B) = \frac{1}{1 + \frac{\alpha_1 B^2}{1 + \frac{\alpha_2 B^2}{1 + \alpha_3 |B|}}}
\label{eq:pade}
\end{align}
This expression can be viewed as a Lorentzian with $B$ dependent width leading to slower $B^{-1}$ decay at infinity. 
It has leading even terms $B^2, B^4$ at $B=0$ and leading $B^{-1}$ term at $B \rightarrow \infty$.
The Pad\'e approximant structure ensures that only even powers of $B$ contribute to the expansion near $B=0$, the first odd term for Eq.~(\ref{eq:pade}) is $B^5$, higher order terms in the nested fraction series would give a higher order for the apperance of the first odd power in the expansion around $B=0$.

\begin{figure}[h]
\centerline{
\includegraphics[clip=true,width=9cm]{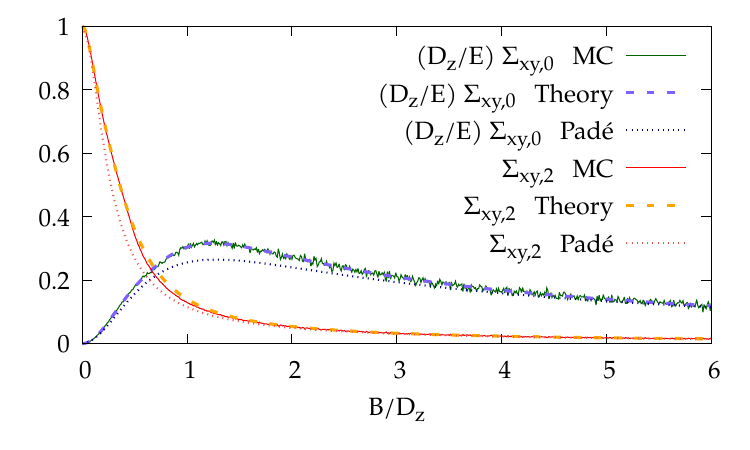} 
}
\caption{The isotropic average of $\Sigma_{xy,0}$ and $\Sigma_{xy,2}$ based on the three methods for $E = 0.1 D_z $.} 
\label{fig2}
\end{figure}

The expressions for $\alpha_{1,2,3}$ are then obtained by matching the leading asymptotic terms around $B=0$ and $B=\infty$ to the power series Eq.~(\ref{eq:avlambda}). Keeping the order $\epsilon^2$ terms gives a simple expression approximation accurate to order $B^4$ at zero and order $B^{-2}$ at $\infty$:
\begin{align}
  \langle \lambda_0 \rangle = u_0 - \frac{4 B^4 (D_z^2 + 3 E^2) u_0}{15 (u_2 - u_0)^3 (u_1-u_0)^3}
  \label{eq:lambda2}
\end{align}
We thus use this expression to fix the values of numerical coefficients $\alpha_{1,2,3}$. The parameters for $\Psi_{z,0}$ and $\Psi_{z,2}$ are essentially fixed by $D_z$ (with $E/D_z$ corrections), while $E$ is the main parameter for $\Psi_{xy,2}$ (with also $E/D_z$ corrections).

Fig.~\ref{fig1} shows the perfect overlap between Monte Carlo averages of $\Psi_{z,n}$ and the summation of the Lagrange inversion series, we used $E = 0.2 D_z$ to show that the expression remains accurate even for relatively large values of $E$. The Pad\'e approximation becomes exact in the limits $B=0,\infty$, thus the worse error for the Pad\'e approximation is 13\% for $\Psi_{z,2}$ at $B/D_z \simeq 1.5$.

\begin{center}
\begin{table}
\begin{tabular}{|c|c|c|c|}
\hline  & $\alpha_1$ & $\alpha_2$ & $\alpha_3$ \\
\hline $\Psi_{z,0}$ & $\frac{1}{D_z^{2}}+\frac{3 E^2}{D_z^{4}}$ & $\frac{0.24}{D_z^{2}} + \frac{1.6 E^2}{D_z^{4}}$ & $\frac{0.6}{D_z} + \frac{2.2 E^{2}}{D_z^{3}} $ \\
\hline $\Psi_{z,2}$ & $\frac{1}{(D_z+E)^{2}}$ & $\frac{4}{5(D_z+E)^{2}}$ & $\frac{1}{D_z}$ \\
\hline $\Sigma_{xy,2}$ & $\frac{1}{6 E^2}+\frac{1}{D_z^2}$ & $\frac{0.28}{E^2} + \frac{-1.23}{D_z^2} $ & $\frac{2.1}{E} - \frac{13.5 E}{D_z^2}$\\
\hline
\end{tabular} 
\caption{Parameters $\alpha_{1,2,3}$ for the Pad\'e approximation Eq.~(\ref{eq:pade}) obtained by asymptotic matching of the Lagrange inversion series up to order $\epsilon^2$ (Eq.~(\ref{eq:lambda2}). }
\label{tabalpha123}
\end{table}
\end{center}

The lineshape $\Sigma_{xy,0}$ is zero at $B=0$ and cannot be approximated by Eq.~(\ref{eq:pade}). To approximate it semi-analytically we introduce the linshape:
\begin{align}
{\tilde P}(B) = \frac{\beta_0 B^2}{(1+\beta_1 |B|)(1 + \beta_2 B^2)} 
\label{eq:Ptilde}
\end{align}
A good approximation to $\frac{D_z}{E} \Psi_{xy,0}$ can be obtained using the values: $\beta_0 = - \frac{4 D_z E}{3 (D_z^2-E^2)^2}$, $\beta_1 = \frac{6}{5 D_z}$ and $\beta_2 = \frac{25 D_z^2}{18(D_z^2-E^2)^2}$ as shown in Fig.~\ref{fig2}.

Fig.~\ref{fig1} also shows the comparison between the Lagrange series and the Monte-Carlo average for $\Sigma_{xy,2}$, we notice that the agreement between Monte-Carlo simulation and Eq.~(\ref{eq:avlambda}) is not perfect and a small discrepancy exists. This is due to the fact that the Lagrange inversion formula is not aware of possible level-crossings between $|\pm 1\rangle$ states and even if the series is convergent it does not necessarily converge to the right root. To account for this we stopped the summation of the Lagrange series at order $\epsilon^4$ for which the best numerical agreement was achieved. The codes for the Lagrange series and Monte Carlo are provided on \cite{tiso_deth}.

These equations Eqs.~(\ref{eq:pade},\ref{eq:Ptilde}) together with Tab.~\ref{tabalpha123} provide a complete set of semi-analytical approximations for the isotropic average of the matrix elements. The functions $\Psi_{z,n}$ are all very similar even if their precise line-shape differs (see Fig.~1), it will thus be difficult to distinguish their contributions from experimental magneto-photo luminescence traces. The characteristic magnetic field scale is of order $D$  with small corrections of order $E^2/D$ but the numerical prefactors depend on $n=0,2$. 

For matrix elements of $\Psi_{xy,n}$ the lineshapes for $\Psi_{xy,0}$ and $\Psi_{xy,2}$ are very different. The E term mixes the $|\pm 1\rangle$ states competing with their mixing by the magnetic field, $\Psi_2^{xy}$ is thus a narrow peak of the same Pad\'e form as  $\Psi_{z,n}$ but with a width mainly fixed by the $E$ value. The mixing of $|0\rangle$ with the two other  $|\pm 1\rangle$ states occurs on an energy scale fixed by $D$ but the small anisotropy breaks $x$, $y$ symmetry perturbatively, the line shape for this matrix element then looks like the field derivative of lineshapes $\Psi_{xy,n}$ with a maximum amplitude of order $E/D_z$. 

Using the relation between the symbols $\Psi$ and matrix elements we find that the lineshape for the matrix overlaps $\mean{|\langle \psi_n |Z \rangle|}$ ($n=0,1,2$ numbers the triplet eigenstates)  is well described by peaks with of order $D_z$, the semi-analytical parameters are given in Table. 2. For $D_z > 0$, the lineshapes for  $\mean{|\langle \psi_0 |X, Y \rangle|}$ will be a combination of  $\mean{ |\langle \psi_0|Z \rangle|}$ and a derivative looking lineshape. Finally matrix elements $\mean{|\langle \psi_{1,2}|X, Y \rangle|}$ will be a combination of lineshapes with a width of order $D_z$ and narrower peaks with width $E$. The case of $D_z < 0$ can be reduced to $D_z>0$ by flipping the order of the eigenstates $n=0,1,2$.

The calculations were performed for each of the eigenstates separately, which is the case of a strong optical polarization of the spin-states. For a spin-distribution close to thermal equilibrium we may need the thermal averages at temperature $T$.
\begin{align}
  \Sigma_z &= \mean{\frac{1}{Z_T} \sum_n  e^{-\lambda_n/T} \langle \psi_n | {\hat S}_z^2 | \psi_n \rangle}
\label{eqTemp}
\end{align}
where $Z_T= \sum_n  e^{-\lambda_n/T}$ is the normalization. We show numerically that it is justified to neglect correlations in this average, and thus the finite temperature average can be  reduced to quantities that we have already computed:
\begin{align}
\Sigma_z &\simeq \frac{1}{Z_T} \sum_n e^{-\mean{\lambda_n}/T} \Sigma_{z,n} 
\label{eqTempU}
\end{align}

\begin{figure}[h]
\centerline{
\includegraphics[clip=true,width=9cm]{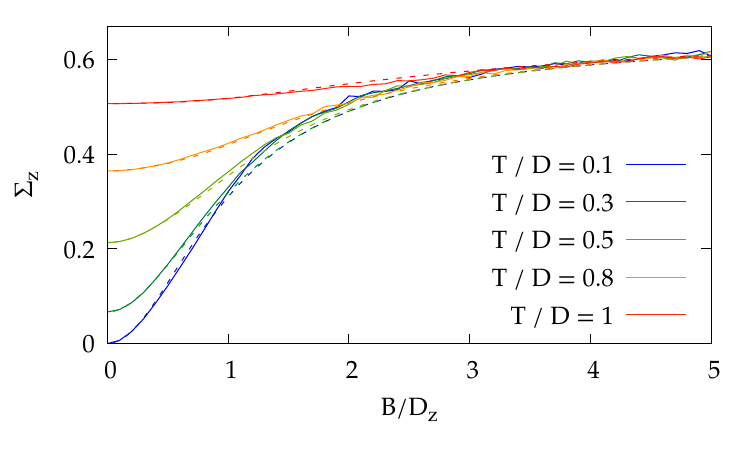} 
}
\caption{Monte Carlo simulation of $\Sigma_z$ the finite temperature averages of the matrix elements for ${\hat S_z}^2$ introduced in Eq.~(\ref{eqTemp}) compared with the uncorrelated approximation Eq.~(\ref{eqTempU}). A very good agreement is observed at both low and high temperatures.
}
\label{fig3}
\end{figure}

The comparison between expressions Eq.~(\ref{eqTemp},\ref{eqTempU}) is shown in Fig.~\ref{fig3}, to approximate $\Sigma_{z,n}$ we use Pad\'e approximation Eq.~(\ref{eq:pade}) (with Table \ref{tabalpha123}) and Eq.~(\ref{eq:lambda2}) for $\mean{\lambda_n}$. We see the uncorrelated approximation gives the finite temperature averages with good accuracy. We notice that in practice the small difference between the Pad\'e approximation and exact lineshapes is likely to be absorbed in a small renormalization of $D_z$ of a few percent. If a systematic high temperature expansion is needed we provide in appendix an identity on the trace of powers of the triplet Hamiltonian that can be useful. It connects the trace of powers of the spin Hamiltonian for $E=0$ with combinatorics.

\section{Conclusion}

In this work, we proposed a theoretical method to find analytically the average over magnetic field orientation for eigenvalues and matrix elements expected to appear in the photo-luminescence rate for triplet excitons. We consider the average characteristic polynomial of the spin Hamiltonian and perform an expansion in powers of the orientation-dependent correction terms, allowing to find the analytical order by order averages. We showed the good agreement between analytical series expansion and Monte-Carlo simulations, and derived semi-analytical approximations for the lines-shapes that we expect to appear in experiments. Since the fine-structure parameters $D_z$ and $E$ can be determined from magnetic resonance experiments with high precision, this analytical  treatment should allow to link experimental MPL lineshape with spin-dependent recombination rates and triplet state populations. 

{\bf Acknowledgments}

The authors thank R. Chowdhury, R.H. Friend and G. Poly for helpful discussions. This research was supported by funding from ANR-20-CE92-0041 (MARS), and the European Research Council
(ERC) under the European Union’s Horizon 2020 research and innovation program (grant Ballistop agreement no. 833350).

\appendix

\section{High temperature expansion}

At finite temperature, the average will involve the trace ${\rm Tr}\;e^{-\beta {\hat H}}$ 
\begin{align}
\Psi(D_z, B) = \frac{2}{3} - \frac{1}{\beta} \partial_{D_z} \langle \log {\rm Tr}\; e^{-\beta {\hat H}} \rangle 
\end{align}
Expansion of the exponential as function of $\beta$ leads to the trace of successive powers of ${\hat H}$.

For the trace of powers of ${\hat H} = D_z {\hat S_z}^2 + B_x {\hat S}_x + B_z {\hat S}_z$, we find the following expansion:
\begin{align}
{\rm Tr} \; {\hat H}^N &= \sum_{2n+2m+p=N} T_{n,m,p} B_x^{2n} B_z^{2m} D_z^p
\end{align}
the coefficients are given by:
\begin{align}
  T_{n,m,p} = \frac{(2m+n)N}{(m+n)(N-n)} \binom{n+m}{m} \binom{N-n}{n+2m}
  \label{eq:Tnmp}
\end{align}
and the sums runs over all integers $n,m,p \ge 0$ obeying the summing condition.

This expresion is obtained by noticing that all the traces of the form 
\begin{align}
{\rm Tr} {\hat S}_{1} ... {\hat S}_{k} 
\end{align}
where each matrix $S_{i}$ is either $S_x$ or $S_z$ are all equal to 0 or 1, Eq.~(\ref{eq:Tnmp}) is then obtained by counting the number of non zero configurations \cite{witschel1971traces}.

\bibliography{detH.bib}

\end{document}